\def\mincir{\raise -2.truept\hbox{\rlap{\hbox{$\sim$}}\raise5.truept
\hbox{$<$}\ }}
\def\magcir{\raise -2.truept\hbox{\rlap{\hbox{$\sim$}}\raise5.truept
\hbox{$>$}\ }}

\documentstyle[12pt,aasms4]{article}

\lefthead{Tr\`evese et al.}
\righthead{Continuum Variability of AGNs}

\begin{document}

\title{Continuum Variability of Active Galactic Nuclei\\
 in the Optical-UV Range}


\author{Dario Tr\`evese\altaffilmark{1}}
\affil{Dipartimento di Fisica, Universit\`a di Roma ``La Sapienza'',\\
Piazzale A. Moro 2, I-00185 Roma (Italy), trevese@uniroma1.it}

\author{Richard G. Kron\altaffilmark{1}}
\affil{Fermi National Laboratory, MS 127, Box 500,\\ Batavia, IL 60510,
rich@oddjob.uchicago.edu}

\and

\author{Alessandro Bunone}
\affil{Dipartimento di Fisica, Universit\`a di Roma ``La Sapienza'',\\
Piazzale A. Moro 2, I-00185 Roma (Italy)}

\altaffiltext{1}{Visiting Astronomer, Kitt Peak National Observatory, 
National Optical Astronomy Observatories, which is operated by the
Association of Universities for Research in Astronomy, Incorporated,
under cooperative agreement with the National Science Foundation}

\begin{abstract}
The variability of the continuum spectral energy distribution has been
analyzed for a complete magnitude-limited sample of quasars in 
Selected Area 57, observed at two epochs in the photographic  $U$, $B_J$, $F$ 
and $N$ bands with the Mayall 4m telescope at Kitt Peak.  
Changes $\delta \alpha$ of the spectral slope $\alpha$ appear correlated with 
brightness variations $\delta log f_{\nu}$ indicating an average hardening of 
the spectrum in the bright phases. This confirms that the correlation of 
variability with redshift, found in a single observing band, is due to 
intrinsic spectral changes.
The average observed 
$\delta \alpha - \delta log f_{\nu}$
relation is consistent with 
the spectral change due to temperature variation of a black body of about
$2.5 \cdot 10^4 K$. 
\end{abstract}

\keywords{galaxies: active - galaxies: multicolor photometry - 
galaxies: Seyfert - photometry : variability - quasars: general}

\section{Introduction}
Variability provides a powerful tool for constraining the physics of
Active Galactic Nuclei.  The original models based on a central black
hole surrounded by an accretion disk (\cite{ree84}) were based on the
constraints on the source size and energy density provided by the
early single-band variability studies.
Single-epoch multi-wavelength observations, covering simultaneously
the range from radio to gamma-ray frequencies, are of crucial
importance in suggesting the physical mechanisms responsible for the
emission. However, in general they cannot sufficiently constrain the
large number of parameters involved in models.
So far multi-wavelength monitoring, with adequate time sampling and duration,
has been possible for a small
number of low-redshift objects, thanks to large international
collaborations (see the review of \cite{umu97} and refs. therein).  
These observations provided further clues to and constraints on the
radiation processes and reprocessing mechanisms, since 
changes of the spectral energy distribution (SED)
associated with flux variations can be interpreted in terms of an
interplay of emission components of different spectral shape and
variability properties.
Various classes of active galactic nuclei (AGN) show different
variability properties. Blazars show strong ($> 1~ mag$) variations of
the observed fluxes on time scales from days to months from radio to
gamma frequencies, while in the optical-UV Seyfert 1 and normal
quasars (QSOs) show smaller ($< 0.5~mag$) variability on time scales
greater than a few months, although $> 1~mag$ variations on time scales
of days have been detected in X-rays in some Seyfert galaxies
(Mushotzki et al. 1993 and refs. therein) (in the following we ignore
the traditional distinction at $M_B$=-23 between Seyfert 1 galaxies
and QSOs).
The result of multi-frequency monitoring indicates that
the optical-UV continuum of AGNs hardens during the bright phases
(\cite{cut85,ede90,kin91,pal94}) in the near infrared-optical-UV range,
while little or no correlation is thought to be present in 
BL Lac objects(\cite{ede92}, see however \cite{mas96a}).

Optical observations at faint magnitudes ($B\approx20 - 23~mag$) enable
magnitude-limited statistical samples to be constructed
(mostly of radio-quiet objects)
in a single field of a large
telescope, and repeated broad-band imaging provides light curves for
each AGN in the sample 
(\cite{haw83,koo88,tre89}(T89), \cite{tre94}(T94), \cite{cri90,haw93,haw96,ber98}(BTK98))
allowing the ensemble variability properties of the sample to be assessed.

Studies of this kind are complicated by the fact that in all magnitude-limited samples
the luminosity is strongly correlated with redshift (L-z correlation), since the majority
of objects lie close to the limiting flux. This makes it difficult to isolate
the intrinsic variability-luminosity (v-L) and variability-redshift (v-z) correlations.

Since early studies, the relation between absolute optical luminosity
and optical variability has been addressed by several authors (\cite{ang72,umo76,pic83}), 
with the aim of discovering whether QSOs
are made up of multiple sources or single coherent sources.
In the former case, a decrease of variability amplitude with luminosity is expected.
In recent years, consensus has progressively grown that brighter QSOs are
less variable on average (\cite{tre94}(T94),\cite{hoo94,cri96})
(see \cite{giv99} for a recent summary). This would seem to support
a model where the active nucleus is powered by a series of supernova explosions
\cite{are94,are97}, though this model can fail to reproduce the variability amplitude in the 
case of most luminous QSOs, according to \cite{haw00}.
A similar negative correlation of X-ray variability with the 1-10 keV luminosity
was found in Seyfert galaxies and QSOs and interpreted in terms of large number of 
incoherent flaring subunits by \cite{gre93}. 
On the contrary, a positive v-L correlation was found by \cite{ede92}
for the Blazar population.

Concerning the L-z correlation,
the situation is further complicated by the fact that the
different variability indicators, used by different authors,  are affected in different ways
by the v-L and L-z correlation and by spurious
redshift dependence due to the combination of time dilation,
intrinsic variability time-scales, and total duration of the observational campaign.
This led Giallongo, Tr\`evese \& Vagnetti (1991) (GTV) to introduce a variability
indicator based on the rest-frame structure function. 
Adopting this indicator, they found
a positive correlation of variability
with redshift, taking properly into account the effect of v-L and L-z correlations.
This result has been subsequently  confirmed by the analysis of the 
QSO structure function in bins of luminosity and redshift,
performed by \cite{cri96} using all the statistical samples available at
that time. 

GTV  suggested that
the v-z correlation can by
explained by a hardening of the spectrum in the bright phases (and vice versa), 
coupled with the increase of the rest-frame frequency with redshift, for a
fixed observing band.  

\cite{dic96}, on the basis of the analysis of various QSO samples,
have shown that indeed, on average, variability increases 
with the rest-frame frequency.
This provides an indirect statistical argument for a
slope change of the quasar SED associated with luminosity variability.
Moreover the increase of variability with rest-frame frequency is
quantitatively consistent with the v-z correlation.

However, a direct statistical quantification of the change of spectral slope
associated with luminosity variation is still missing.
Moreover, even though variability has historically played a key role in
constraining models of the QSO central engine,
the physics of variability is still largely unknown.
In fact, quite diverse variability mechanisms
have been proposed, including supernovae explosions
(\cite{are94}), instabilities in the accretion disk (\cite{kaw98}), and gravitational 
lensing due to intervening matter (\cite{haw93,haw00}). 
Discriminating between these different models on the sole basis of single band correlation
functions and v-L and v-z correlations is a very difficult task.

In the present paper we analyze $UB_JFN$ photometry of the faint QSO
sample of SA57 to look for direct evidence of an average spectral hardening 
for increasing  flux. 
A statistical quantification of the spectral slope changes as a function
of luminosity variations provides a new constraint for models of the 
variability mechanism.
We also discuss  the consistency of the
SED variations with the v-z correlation and
with temperature changes of an emitting black body.

This paper is organized as follows: data and reduction procedures are
described in section 2; the statistical analysis of SED changes and
their relation with flux variations are described in section 3; a
concluding discussion is presented in section 4.

\section{The AGN Sample and the Spectral Energy Distributions}

Statistical studies of variability require a QSO sample 
based on clearly known selection criteria, whose  
effects are clearly understood and quantifiable.
Among the existing QSO samples, the magnitude-limited sample of faint QSOs of Selected Area 57 
(\cite{kro81,kkc}(KKC)) is probably the most studied.
This field has been repeatedly observed since 1974 with the
Mayall 4 m telescope at Kitt Peak in the photographic $U,B_J,F,N$ bands.
Different techniques, such as colors, lack of proper motion, and
variability, have been applied to search for AGNs down to $B_J=23$
and to estimate the completeness of the sample. Spectroscopic confirmation of
almost all of the QSO candidates has been obtained
(\cite{kro81,kkc}(KKC),\cite{koo88},T89,\cite{maj91},T94),
with the exception of the faint part of the sample of extended sources of BTK98.
This allows for the first time to perform a statistical analysis of the
variability of the spectral energy distribution of QSOs.
Single-band variability studies (T89, T94, BTK98) of the AGN sample where
based on the entire collection of J (or $B_J$) plates. 
However, as discussed in
the next section, the most reliable analysis of 
SED variability is obtained considering only
``simultaneous'' $U,B_J,F,N$ observations. For this reason we
selected the two sets of plates taken  in April 1984 and in April 1985.
The list of the plates  is given in Table 1, which contains 
plate identification, the date, and the photometric band.
The effective bandwidths of the $U,B_J,F,N$ bands are 
570, 1250, 1150, 1720 \AA~ respectively and the effective wavelengths
are 3540, 4460, 5980, 7830 \AA respectively.

\placetable{tbl-1}

Photometric methods and signal-to-noise ratio optimization are described
in T89, T94 and BTK98.
The AGN sample consists of 40 spectroscopically confirmed objects.
Table 2 gives the photometry on each of the 9 plates considered
in the present analysis.

\placetable{tbl-2}

Of these 40 objects,   35 
are  the QSOs discussed in T94, while another 5 objects
are the spectroscopically confirmed
AGNs  selected by BTK as variable objects with extended images.
The limiting magnitude of the sample is $m_2 < 23.5~mag$, where $m_2$ is
the $B_J$ magnitude evaluated in a fixed
aperture of radius $0''.5$ (but see BTK98 for faint extended objects).
The cumulative redshift distribution has 25\%, 50\%, and 75\% of the objects
below z=(0.75, 1.16, 1.80) respectively.

\placefigure{fig1}

Figure 1 shows the Hubble diagram of the sample.
Figure 2 shows a comparison of the $U,B_J,F$ fluxes at one epoch (April 1984),
normalized at $\lambda=2000$~\AA, as a function of the rest-frame 
frequency. 
\placefigure{fig2}
Most objects have similar spectral slopes, with five main exceptions,
which are objects with a much steeper spectrum.
Two of them are low redshift and two are the highest redshift objects.
The  former two, N$_{ser}$ 110195 with z=0.243  and N$_{ser}$ 114264 with 
z=0.287, belong to
the sample of BTK98 and are relatively faint AGNs, with $M_{B_J}=-19.6$
and $M_{B_J}=-20.7$ respectively. 
Their r.m.s. variability is among the lowest
in the sample of BTK. 
Since we know that, in general, fainter AGNs show a stronger variability,
we may argue that in this case the light from the nucleus is diluted
by the stellar light of the host galaxy.
This could  also explain the steep
spectrum. If so, the spectral changes would also be affected by the presence
of a constant stellar component.
Two other steep-spectrum objects are the most distant in the sample,
N$_{ser}$ 111610 with z=3.02  and N$_{ser}$ 101392 with z=3.08, 
and $M_{B_J}=-20.7$ and $M_{B_J}=-24.9$
respectively (bottom right in Figure 2).
Their spectral slopes are strongly affected by  intergalactic
Ly-$\alpha$ absorption. In particular the $B_J$ band falls in the (rest-frame)
region between the emission Ly-$\alpha$ and the Lyman-limit, and
the $U$ band falls beyond the Lyman-limit.
The relevant  fractional attenuation of the continuum
can be evaluated as $\magcir$ 0.2 and
and $\magcir$ 0.5 respectively, e.g. from \cite{ste87}.
Notice that, in Figure 2, the  downwards shift of  the F-band flux is due 
to the normalization at 2000 \AA~ computed from the extrapolation of the
steep uncorrected spectrum, and is dramatically reduced if the extrapolation
to 2000~ \AA~ is done after the correction for the Ly-$\alpha$ absorption.
Once corrected for the absorption the two spectra become consistent with 
the average slope $\langle \alpha \rangle \approx -0.5$
of the sample (see \cite{gia90}).
We stress that the changes of the spectral index due to variability
are independent of the 
intergalactic absorption, even though the apparent SED is affected.
Figure 3 shows, with arbitrary scale, the data reported in Table 2.
\placefigure{fig3}

In the majority of the cases the deviations of the SED from a linear 
$log f_{\nu}- log \nu$ relation are small, but some objects 
have a local maximum of emission in the observed frequency range.
Of course, such a humped SED can be relevant  for physical models
of the emission mechanism.
Correlations of colors with other
properties, e.g. the X-optical slope (\cite{miy97}),
can be affected by artificial curvature due to non-simultaneous measures 
of different bands.

\section{Variability of the continuum spectral energy distribution}

Our  aim is to correlate the SED changes with the brightness variations.
We do this by correlating the variations $\delta \alpha$ of the spectral slope 
$\alpha$,
defined by $log f_{\nu} \propto \nu^{\alpha}$, with the variation of 
$\delta log f_{\nu}$, where $\nu$ is the rest-frame frequency of one
of the observing bands.
In order to minimize the effect of noise, we are subject to several 
restrictions.
On average  the best signal-to-noise ratio is obtained on $J$ plates;
thus we use pairs of $J$ plates to compute $\delta log f_{\nu_J}$.
The noise in $\alpha$ is lower for larger leverage on 
the $log \nu$
axis and the larger the number of bands used. The data used for computing
$\delta log f_{\nu_J}$ must not be used in the calculation of $\alpha$,
otherwise a spurious correlation is found, namely $\delta \alpha$ and 
$\delta log f_{\nu_J}$ appear correlated even in the absence of variability,
due to the presence of correlated noise in both $\delta \alpha$ and 
$\delta log f_{\nu_J}$. This statistical bias may even lead to  strong 
and highly significant correlation (\cite{mas96}).
It must be considered also that the SED cannot be represented exactly 
by a power law in the wavelength range studied. 
This is not a problem as long as
$\alpha$ at different epochs is computed in a similar way.
In fact $\alpha$ is a linear combination of the
values of $log f_{\nu_J}$ measured in different bands, thus any deviation
of the SED from a power law, while affecting $\alpha$,
does not affect  $\delta \alpha$, which is a linear combination of 
$\delta log f_{\nu_J}$ values. This is true even when
the deviation from the power law is produced by a (non-variable)
absorption in some of the bands.
However, computing $\alpha$ in a very different way at two epochs,
e.g. using $U$ and $N$ or $J$ and $F$ bands,
can generate $\delta \alpha$ values not due to variability.
This 'spurious' $\delta \alpha$ can even be different for two
identical QSOs, if they have different redshifts.
The above considerations led us to select the two-epoch set
of plates described in 
Table 1, to minimize spurious effects and
to increase the reliability and robustness of the results.
We  determined $\alpha$ from a linear fit
in the $log f_{\nu}-log \nu$ plane from $U,B_J,F,N$  data at the first 
epoch and from $U,F,N$ data at the second epoch (the $N$ magnitude
of N$_{ser}$ 104855 is missing at the first epoch). In this way two more
$J$ plates (MPF3919 and MPF3977), one at the first and one at the second epoch, 
are available to compute an independent $\delta f_{\nu_J}$.
Emission lines, which are known to vary on the same time scale of the 
continuum emission, may also affect the spectral slope variations. 
Whether they cause a positive or negative contribution
to the slope depends on the particular band and redshift.
Of course, if line variation were perfectly synchronous and proportional to continuum variation
they would not cause any additional slope variability.
A recent paper of \cite{kas00}, provides light curves of $H_{\alpha},H_{\beta},H_{\gamma}$ lines
and continuum of 28 PG QSOs with a sampling interval of 1-4 months for a total
observing period of 7.5 years.
From the line-continuum cross-correlation function the authors derive time delays 
of the order of 100 days for 17 of the 28 QSOs. 
$H_{\alpha}$ flux fluctuations are smaller than 20\% of the continuum fluctuations.
Continuum and line variation are, in general,
strongly correlated  and the time width of the cross correlation function,
which depends on the power spectrum of variability, is larger than the delay.
As a result, in most cases the cross-correlation at zero time-lag is large ($\magcir 0.5$)
so that line variations synchronous and proportional to the continuum variation can be considered 
a good approximation for the purpose of the present analysis.

Although we cannot correlate the slope changes with flux variations for each
individual object, we can study the ``ensemble average'' of
spectral variations of the 40 objects of our sample between two epochs.
In Figure 4, $\delta \alpha$ versus $\delta log f_{\nu_J}$ is reported
for the entire sample, with the two linear regression lines, 
showing an average increase of the spectral slope $\alpha$
for increasing $f_J$, i.e.  a hardening of the spectrum in the
brighter phases. 
The distribution of $\delta \alpha$'s is symmetric about zero, indicating 
the absence of any strong systematic effect on the spectral slopes.
\placefigure{fig4}
The correlation coefficient between $\delta \alpha$ and $\delta log f_{\nu_J}$
is $\rho=0.39$ and the
probability of the null hypothesis is $P(>\rho)=1.28 \cdot 10^{-2}$.
We take the above result as  statistical evidence for a relation of the
type $\delta \alpha =a + b~\delta log f_{\nu_J}$ between $\alpha$ 
and luminosity
variations. 
To check the dependence of the result on the two most deviant points
($N_{ser}$=118122 $\Delta log f_J$=-0.276, and $N_{ser}$=104885 
$\Delta log f_J$=0.292) we evaluated the correlation excluding in turn: i) both
the points, ii) the first point  and iii) the second point. 
We obtained respectively:
i) $\rho=0.288$ $P(>\rho)=7.98 \cdot 10^{-2}$, 
ii) $\rho=0.290$ $P(>\rho)=7.33 \cdot 10^{-2}$ and
iii) $\rho=0.395$ $P(>\rho)=1.29 \cdot 10^{-2}$.
To determine the slope $b$ of the above relation, the errors on 
$\delta \alpha$ and $\delta log f_{\nu_J}$ must be taken into account,
especially if they differ on the two axes.
Errors on $\delta log f_{\nu}$ for the plate pairs of each band can be 
computed as follows. From the photometric catalog of the area of sky  
which has been monitored (KKC,T89,T94,BTK98), each containing 2540
objects brighter than $B_J$=22.5, we computed $\delta log f_{\nu_J}$
for each object and for each band. Then, in each band, we sorted the objects 
according to 
their magnitude, and for each object we took the 100 nearest neighbors (in magnitude)
and computed the r.m.s. value  $\sigma_{\delta_i}$ of  $\delta log f_{\nu_i}$ 
(i=$U,B_J,F,N$), after the exclusion of the points lying  more than three 
standard deviations from the mean. 
As most of the objects in the field are non-variable
we assume  $\sigma_{\delta_i}$ as a measure of the r.m.s. photometric noise on 
the flux differences $\delta f_{\nu_i}$. 
Using the entire plate set, consisting of 5, 15, 5, 6   plates in the
$U,B_J,F,N$ bands respectively, 
it is also possible to derive a magnitude dependent photometric error 
for each plate (instead of considering the error 
on the magnitude difference in plate pairs). As above we sorted the 2540 
objects according to their $B_J$ magnitude (on some reference plate), then
for each object we considered the light curve, its time average and its mean
square dispersion, in each band.  Then we took 100 neighboring objects in magnitude
and computed the r.m.s.
dispersions around the time average.  Neglecting the
variance of the time average, this provides an estimate of the
r.m.s. photometric error of individual plates.  
The errors estimated from plate pairs are consistent with 
a quadratic combination of the errors of the relevant individual plates.
These errors are shown for each point of Figure 3.

The variance of $\delta \alpha$ is a linear combination of the variances
of the signals in the bands involved, with appropriate 
coefficients depending on the frequencies of the bands.
We can fit to the $\delta \alpha$,$\delta log f_{\nu_J}$ data
a straight line taking into account the errors of individual points on both 
axes. The actual calculation is done using the ``fitexy'' subroutine
from (\cite{pre92}), which gives also the statistical 
uncertainties on $a$ and $b$. The result, with one sigma errors, 
 is $a=(-8.49 \pm5.50) 10^{-2}$ and $b=(2.55 \pm 0.75)$.

Repeating the analysis with the exclusion of the N band reduces the effect of a 
deviation of the SED from a power low providing a less biased, but more noisy,
estimate of the local spectral slope. The  correlation is
lower, but still positive and marginally significant 
($\rho=0.27, P(>\rho)=8.87 \cdot 10^{-2}$).
The straight line fit to the data, $a=(-1.67 \pm 3.07) 10^{-3}$ and
$b=2.29 \pm 0.37$, is quite consistent with the previous result. 

\cite{dic96} found that the average increase of variability with redshift
is $\Delta S_1/ \Delta \log (1+z) \simeq 0.25-0.30$.
This quantity is related to the slope $b$ by:
 $\Delta S_1/ \Delta \log (1+z) \simeq \Delta S_1 /\Delta \log \nu \simeq
\langle \delta \alpha ^2 \rangle^{1/2} \simeq b~\langle (\delta \log f_{\nu})^2 \rangle^{1/2}$,
where  $\langle \delta \alpha ^2 \rangle^{1/2}$ is the r.m.s. fluctuation of the 
spectral slope and $S_1$ is the variability indicator defined in \cite{dic96} and 
$\langle (\delta \log f_{\nu})^2 \rangle^{1/2} \simeq 0.1$ is the r.m.s. flux density
fluctuation. Thus the value of $b$  found in the present analysis is consistent
with the increase of variability with frequency found by \cite{dic96}
and confirms  the interpretation of the positive v-z correlation suggested by
GTV , by a direct observation of the r.m.s. slope variability.
We can compare our statistical results with the 
$\alpha-log f_{\nu}$ relation found by  \cite{ede90} in 
individual objects
through the analysis of IUE observations. All of their 6 objects
have low redshift and are sampled around a wavelength $\lambda \simeq 2000$ 
\AA~ which is  about  the  average wavelength
of our sample ($<z>=1.365$). For 5 of their objects 
they find a significant correlation of $\alpha$ and $log f_{nu}$. For 4 of 
these objects the slope $b$ of  the relation $\alpha =b~log f_{\nu} + const$
is around 1.7, while 3C 273 shows a steeper ($b\approx 2.8$) relation.  

It would also be interesting to compare our findings with the
variability of (B-R) colors measured by \cite{giv99} in a
sample 42 of PG QSOs. Although the average redshift and the rest-frame
wavelength region sampled, $<z> \simeq 0.12$, $\lambda \approx 3500$ \AA~
differ from ours, qualitatively their results agree with ours,
namely they also find a hardening of the spectrum during bright
phases. 
A more detailed and quantitative comparison requires a re-analysis of
\cite{giv99} data in a consistent way 
(Trevese \& Vagnetti, in preparation; Trevese \& Vagnetti 2000a,b).

\section{Discussion and Summary}

Let us assume that the SEDs of the objects of our sample are typical of QSOs,
namely they are  dominated by the big blue bump in the spectral region
around  $\lambda=2000$ \AA~ (see e.g. \cite{bre90}).
We want to check the hypothesis that both the 
SED slope and brightness changes are caused by temperature changes of an 
emitting black body.
For this purpose we use $U,B_J,F$ data only and we assign the slope $\alpha$ to 
the intermediate frequency $log \overline{\nu}=\frac{1}{3} log (\nu_U \nu_J \nu_F)$.
Figure 5 shows $\alpha$ versus $\log \bar{\nu}$ for each object.
In the same figure the curves $\alpha(x)\equiv 3 - x e^x/(e^x-1)$,
$x\equiv h \nu/kT$, 
of black bodies  are also
reported, for various values of the  temperature T,
h and k being the Planck and the Boltzmann constants. 

\placefigure{fig5}

The two points 
marked with stars represent the uncorrected SED of the two highest redshift
objects, which are affected by the intergalactic Ly-$\alpha$ absorption
and appear particularly steep. Since SA 57 is close to the Galactic pole,
interstellar reddening has a negligible effect on the location of 
points in Figure 5.
The distribution of $\alpha$ is skewed towards negative values,
with most objects around -0.5. 
To check the consistency of slope and brightness variation in a more
stringent way, we exclude from the sample
the most deviant points, with $\alpha < -2 \sigma_{\alpha}$. 
Then we obtain $<\alpha>=-0.48 \pm 0.68$.
The corresponding average 
temperature is $T \approx 25000 K$.
We cannot compare $\delta \alpha / \delta log f_{\bar{\nu}}$ with
$\alpha$ of individual objects because the noise is too large, especially on
the former quantity. However we can compare 
the slope $b$ of the straight line fitting the points in Figure 4,
representing the average increase of the spectral slope $\alpha$
for increasing luminosity,
with  the corresponding relation expected for black body 
spectra of varying temperature.
Brightness and slope variations of the SED of a black body of fixed surface
are related by 
$(d \alpha/dT)/(d log B_{\nu}/ dT)= (\ln 10) [1-x /(e^x-1)] \equiv F_{BB}(x)$.
In Figure 6, $F_{BB}(x)$ and $\alpha(x)$ are plotted as functions of x. 
\placefigure{fig6}
The value of $<\alpha>$ and the r.m.s. spread $\sigma_{\alpha}$ of the sample
define an interval of x in this plot.
In the same figure is also reported the value of $b$ and $(b - \sigma_b)$
as deduced from the statistical analysis of section 2, which define a lower 
limit on x.

As seen from Figure 6, the above values define      
a non-empty region of consistency between $F_{BB}(x)$ and $\alpha(x)$
for the sample.
A black body with $x \approx 3$ satisfies simultaneously the two constraints. 
Thus we can say that the $\delta \alpha$ and $\delta log f_{\nu_J}$,
observed at an average rest-frame $\bar{\lambda} \approx 2000$ \AA~ 
are consistent with temperature variations of a typical black body 
of temperature $T \approx 25000 K$.

A single black-body spectrum is an oversimplified representation of the SED
which, in the case of accretion disks models, ignores temperature gradients and
the presence 
of other components  necessary to explain the emission outside the frequency 
range considered. In any case  we can say that:

i) The increase of the amplitude of variability
with the rest-frame frequency (\cite{gia91,cri96,dic96}) is due
to the  hardening in the bright phases (and vice versa)
of the spectrum, which to a first approximation maintains 
its local power-law shape.

ii) The r.m.s. slope variations are quantitatively consistent
with the increase of variability for increasing frequency found by \cite{dic96}.
This  confirms, by a direct observation of spectral slope changes, that 
correlation between the variability  and redshift is accounted for
by intrinsic spectral changes and does not require  cosmological
evolutionary effects.

iii) Our results show that the average changes of the SED slope and the 
relevant flux variations 
in the optical-UV region are at least not inconsistent with 
temperature variations of a single emitting black body of 
$T \approx 2.5 \cdot 10^4 K$.

iv) The average relation between slope and brightness variations
provides a new constraint on  models of the emission mechanism.

For instance, an optically thin plasma model 
(\cite{bar93}) requires a temperature of the order of $10^6 K$ 
to explain with a single
emission component the SED from optical-UV to X-ray frequencies (see 
\cite{fio95}). The bremsstrahlung luminosity 
$L_{\nu} \propto T^{-1/2} e^x$ (\cite{rib79}), 
can provide  an optical-UV spectral slope $\alpha$ consistent with 
the observations, but in this case the slope and brightness changes produced by
temperature changes are related by
$F_{ff}(x) \equiv \partial \alpha / \partial log f_{\nu} = x/(x-1/2)$,
i.e. $F_{ff}  \simeq -2x$, for $x \ll 1$. 
Thus, for $T\approx 10^6$, $F_{ff} \approx -0.1$ is not only
too small in absolute value, but it is even negative, thus completely
inconsistent with the result of our analysis.
Notice that also the temperature changes of a simple black body of 
$T \approx 10^5 K$  would be inconsistent with our findings
since, for $\lambda \approx 2000$ \AA, this temperature implies
$F_{BB} \simeq 0.27$.
In fact, at high temperature any thermal emission approaches the limit
where the long-wavelength part has a fixed spectral slope, independent of 
temperature changes (\cite{pal94}).
Any  model trying to reproduce the observed SED
from the IR to X-rays could be tested against the observed 
$\alpha- log f_{\nu}$ relation. 
In particular, it is possible to investigate whether the variation of
one, or some, of the parameters involved (like the temperature) are
capable of reproducing the observed relation.  Perhaps additional
phenomena, e.g. hot spots or flares (\cite{kri94,kaw98}) 
over an otherwise stationary accretion disk,
are needed to explain the SED variability.  
The main limit of the present analysis are the small amplitude of variability,
due to the short time interval (1 yr) between the observations, and the 
lack of $\alpha- log f_{\nu}$ relation for individual objects, which could
be different  depending, e.g., on their intrinsic luminosity.

Despite these limits, our results clearly show that a few additional
multi-band observations, properly distributed in time, could provide
strong constraints on the physics of emission and variability,
especially if associated with simultaneous observations in the IR and X-rays.

\acknowledgments

This work was partly supported by the Italian Ministry for University
and Research (MURST) under grant Cofin98-02-32

\clearpage
\figcaption[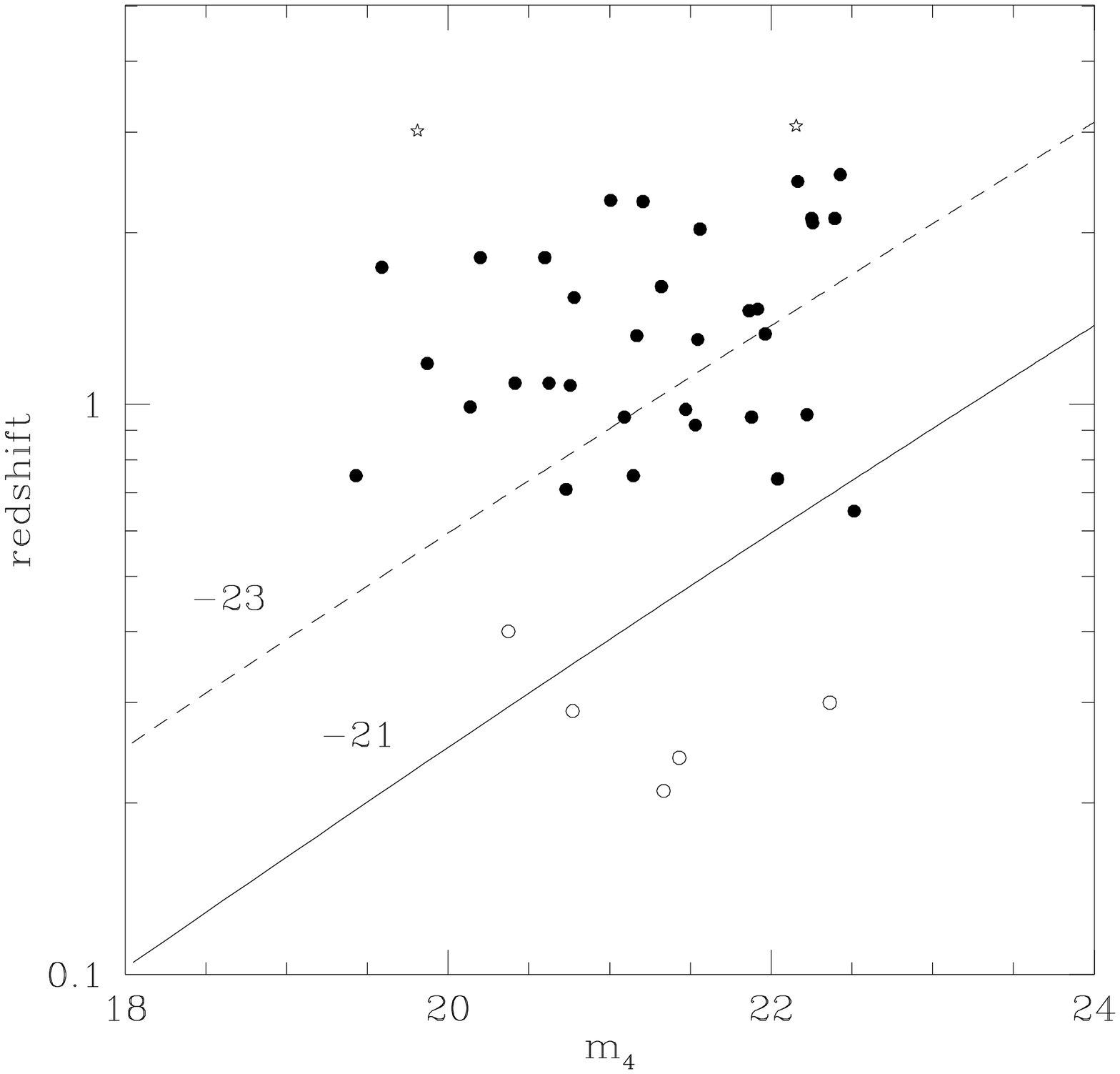]{Redshift of all the objects of the sample vs.
the apparent magnitude ($m_4$), defined as the $B_J$ magnitude 
evaluated in a fixed aperture of  radius of 1''.1. 
The stars represent the two highest
redshift objects. The open circles
represent the five objects with extended images from BTK.
Lines of constant luminosity $M_J=-23$ and $M_J=-21$
are drawn assuming $H_o=~50~km~s^{-1}Mpc^{-1}, q_o=0.5, and \alpha=-1$.
\label{fig1}}

\figcaption[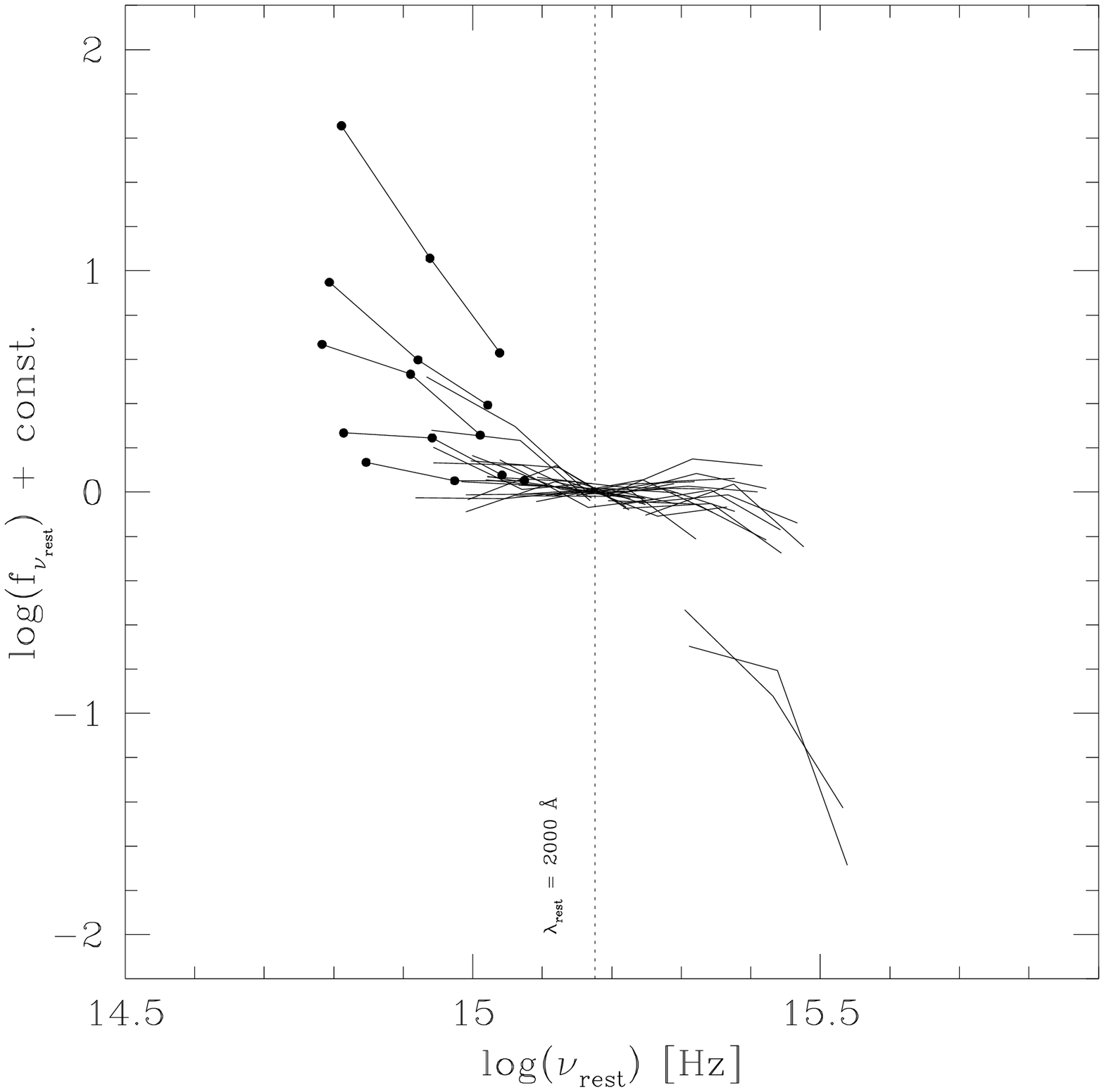]{ $U, B_J, F$ flux as a function of the rest-frame
frequency, normalized at 2000 \AA, for the 40 objects of the sample.
The photometric data of the 5 objects from the list of BTK98 are represented 
by dots. 
\label{fig2}}

\figcaption[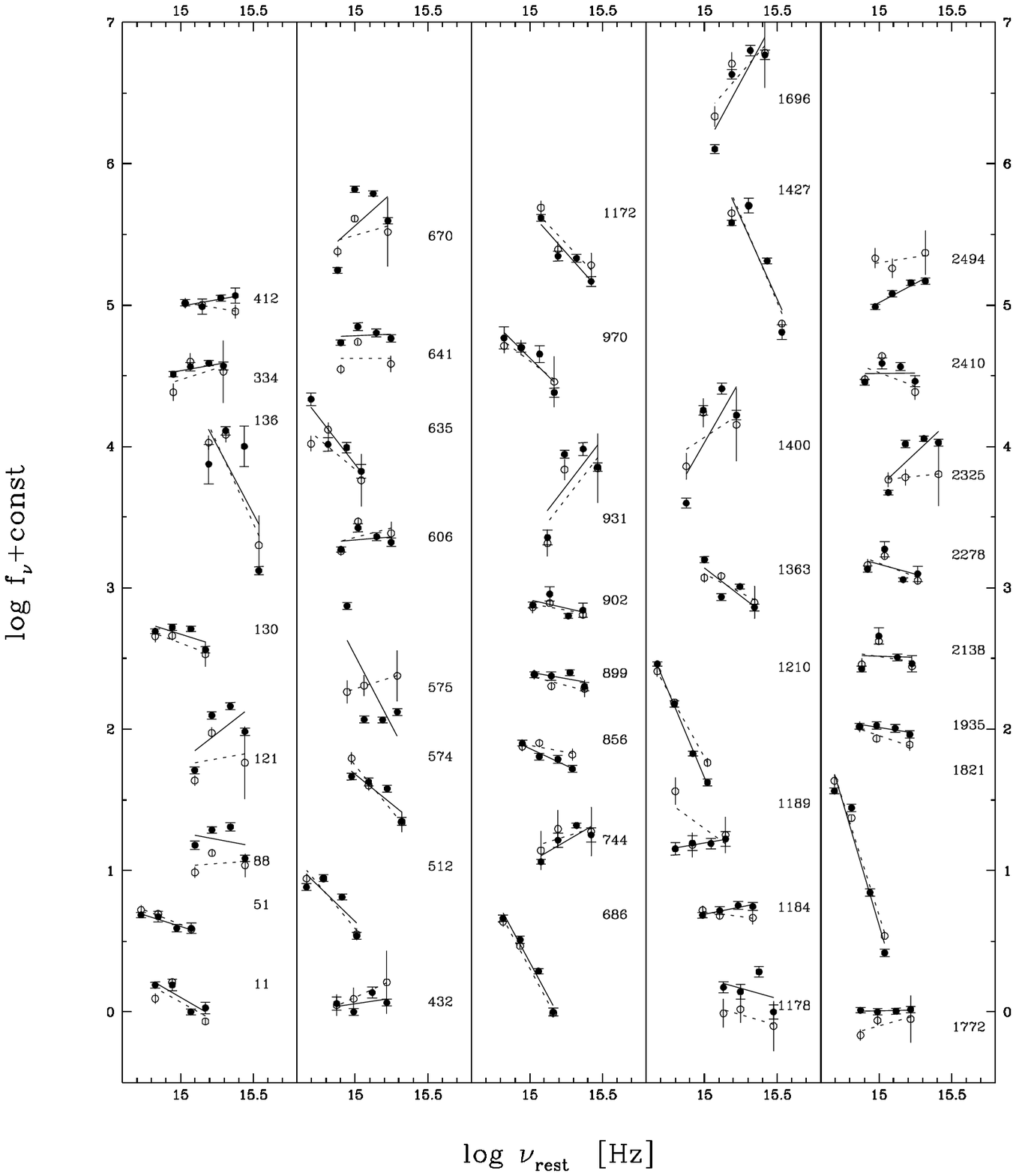]{The spectral energy distribution in arbitrary scale.
{\it Filled circles}: first epoch; {\it open circles}: second epoch.
The relevant power-law fits are also shown. The scale is the same for each 
pair of data sets representing the same objects at two different epochs.     
\label{fig3}}

\figcaption[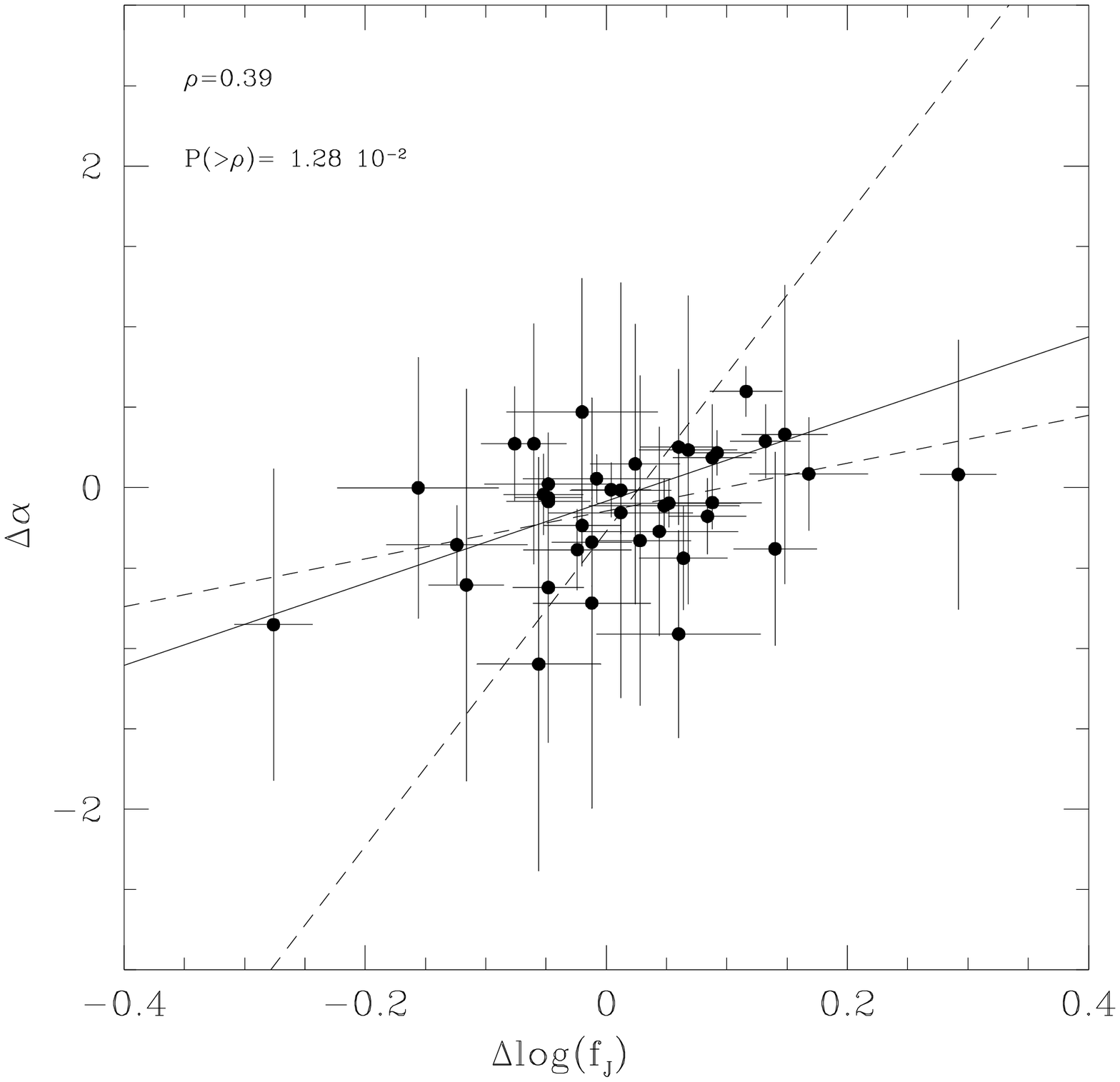]{$\delta \alpha$ versus $\delta log f_{\nu}$,
with $f_{\nu}$  measured in the $B_J$ band, on independent plates (see text).
The dashed lines represent the linear regressions, from which the
correlation coefficient $\rho$ is computed. The solid line is the linear 
fit which takes into the errors in both coordinates. 
\label{fig4}}

\figcaption[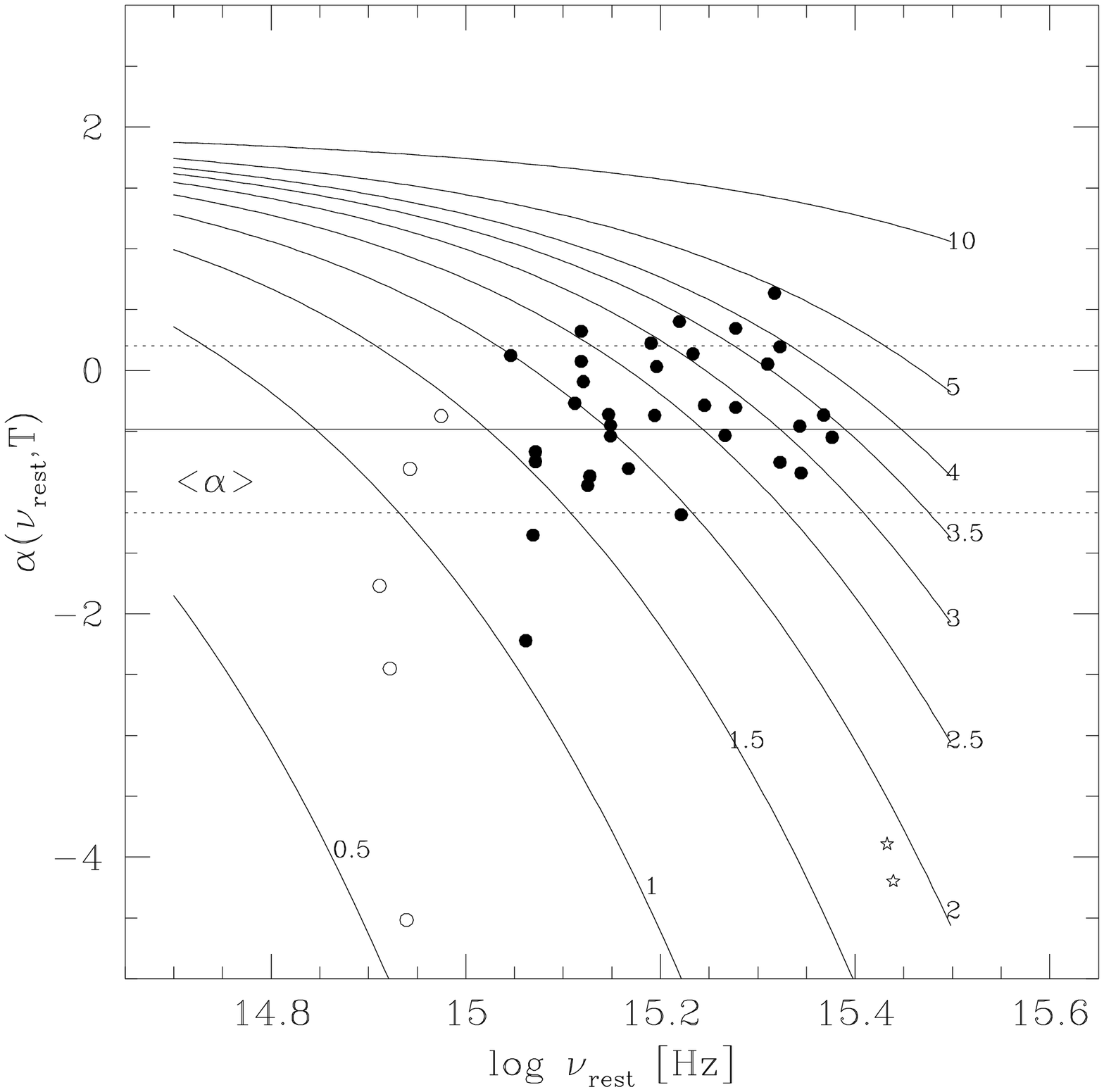]{The slope $\alpha$ as a function of
$x \equiv h \nu/kT$,  for the 40 objects of the sample, as computed
from $U, B_J, F$ bands, with $log \nu =\frac{1}{3} log (\nu_U \nu_B \nu_F)$. 
The stars represent the two highest redshift objects whose colors are 
affected by Ly$_{\alpha}$ absorption. The open circles
represent the five objects with extended images from BTK.
The solid lines represent $\alpha(x)$ for black bodies of
different temperatures T, indicated in units of $10^4$ K. 
The horizontal lines represent $<\alpha> \pm \sigma_{\alpha}$, after
the exclusion
of the 3 points at more than 2-sigma from the mean. 
\label{fig5}}

\figcaption[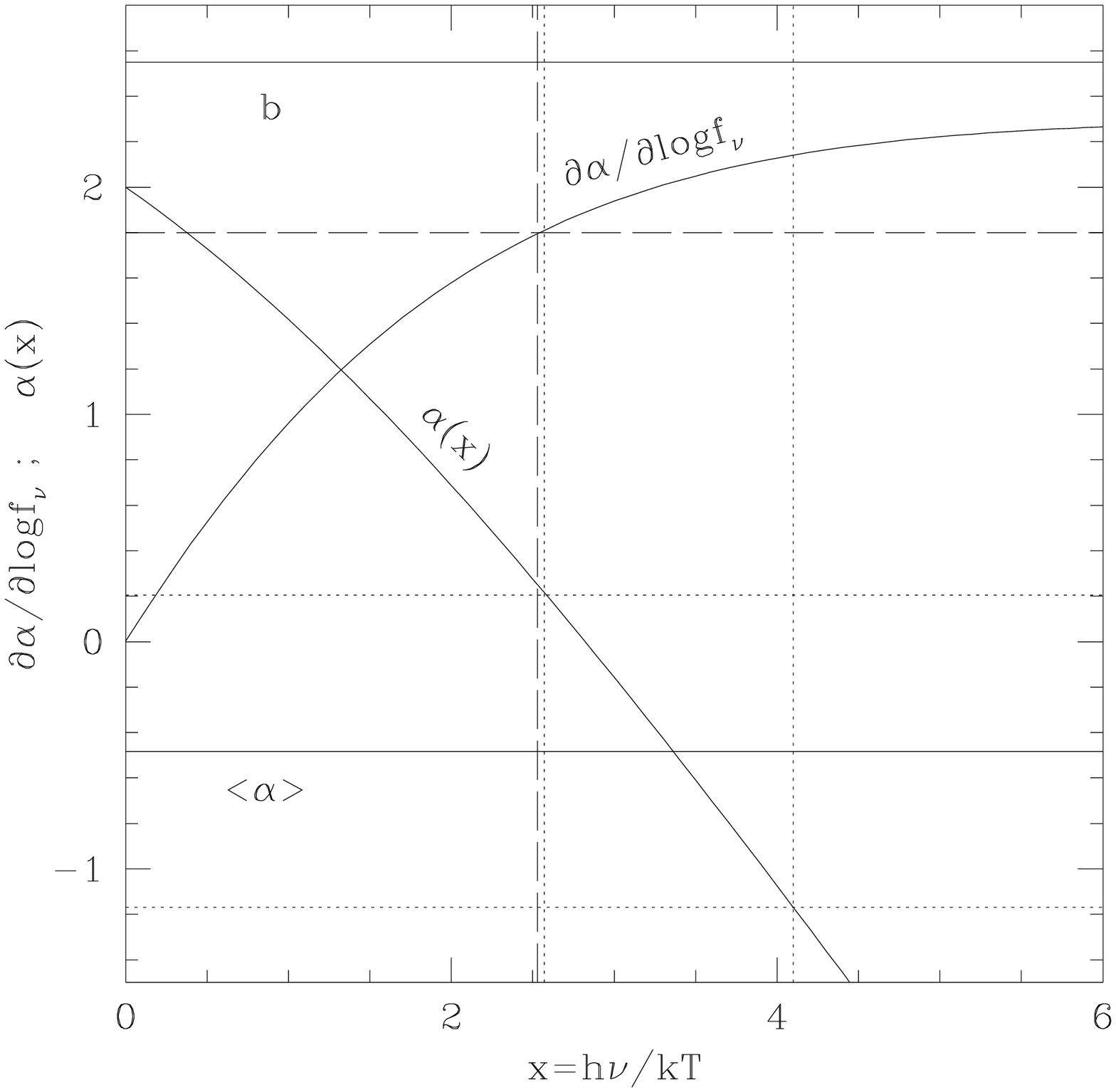]{$\alpha(x)$ and $F_{BB}(x)$ for a black body, as a 
function of $x \equiv h \nu/kT$. The continuous horizontal lines
represent respectively $<\alpha>$ and the slope b of the straight
line fitting the data in Figure 3. The dashed and dotted lines 
represent the one-sigma interval and the relevant bound on x, 
deriving from the comparison of the data with the black body curves
$F_{BB}(x)$ and $\alpha(x)$ respectively. 
\label{fig6}}

\clearpage

\begin{deluxetable}{ccc}
\footnotesize
\tablecaption{Plate Journal. \label{tbl-1}}
\tablewidth{0pt}
\tablehead{
\colhead{MPF \#}  & \colhead{UT date} & \colhead{Band}
}
\startdata
3919 & 1984~Apr~05 & $B_J$ \\
3920 & 1984~Apr~05 & U \\
3921 & 1984~Apr~05 & $B_J$ \\
3922 & 1984~Apr~05 & F \\
3923 & 1984~Apr~05 & N \\
3973 & 1985~Apr~25 & N \\
3975 & 1985~Apr~26 & F \\
3976 & 1985~Apr~25 & U \\
3977 & 1985~Apr~25 & $B_J$ \\
\enddata
\end{deluxetable}

\clearpage
 
\begin{figure}
\plotone{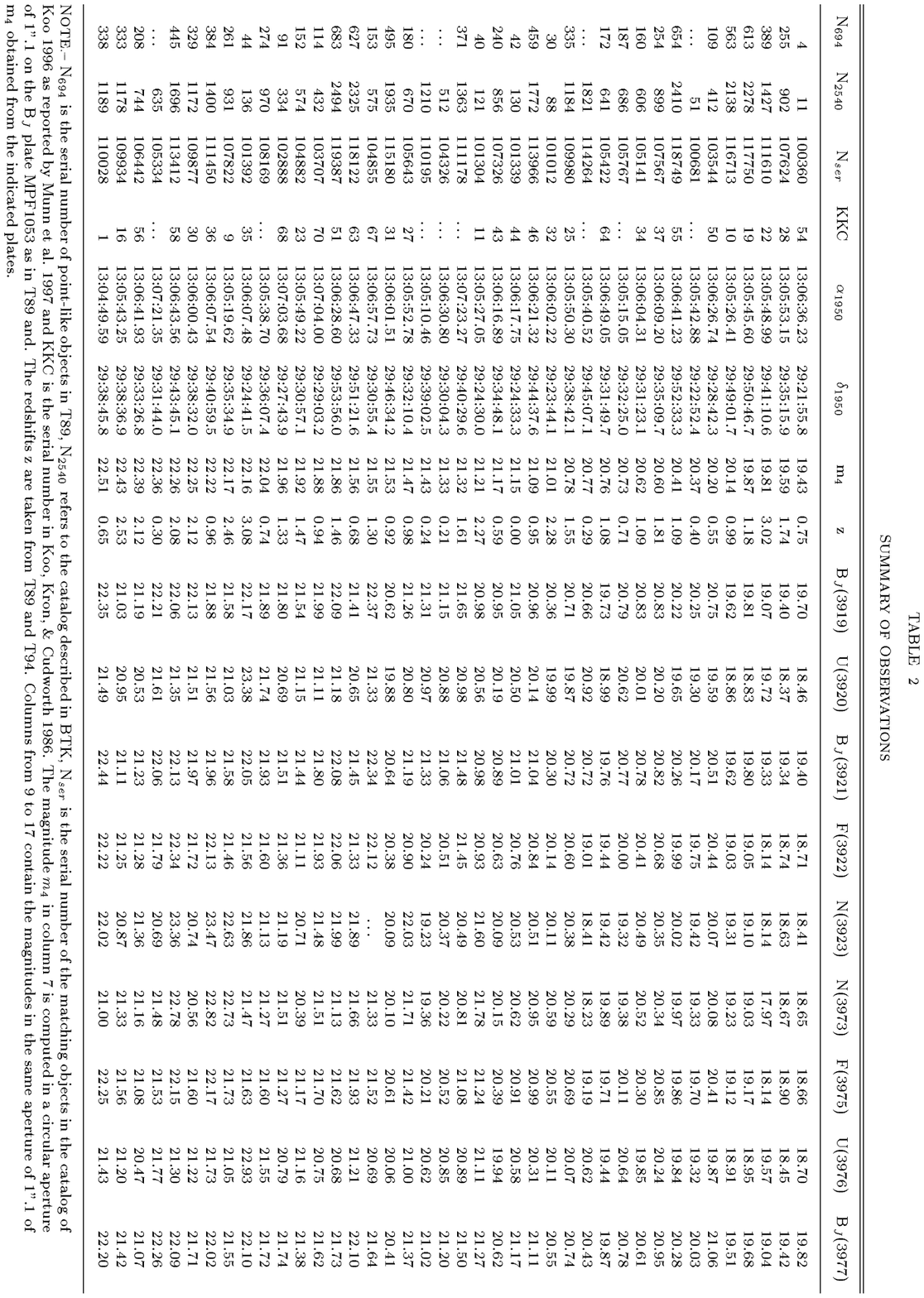}
\vskip 0cm
\end{figure}
\clearpage

\clearpage

\begin{figure}[1]
\plotone{fig1.eps}
\vskip 0cm
\end{figure}
\clearpage

\begin{figure}[2]
\plotone{fig2.eps}
\vskip 0cm
\end{figure}
\clearpage

\begin{figure}[3]
\plotone{fig3.eps}
\vskip 0cm
\end{figure}
\clearpage

\begin{figure}[4]
\plotone{fig4.eps}
\vskip 0cm
\end{figure}
\clearpage

\begin{figure}[5]
\plotone{fig5.eps}
\vskip 0cm
\end{figure}
\clearpage

\begin{figure}[6]
\plotone{fig6.eps}
\vskip 0cm
\end{figure}
\clearpage

\end{document}